# The Requirement for Cognition, in an Equation

Robert Worden

Theoretical Neurobiology Group, University College London, London, United Kingdom

rpworden@me.com

Draft 0.9; May 2024

Abstract:

A model of the evolution of cognition is used to derive a Requirement Equation (RE), which defines what computations the fittest possible brain must make – or must choose actions as if it had made those computations. The terms in the RE depend on factors outside an animal's brain, which can be modelled without making assumptions about how the brain works, from knowledge of the animal's habitat and biology.

In simple domains where the choices of actions have small information content, it may not be necessary to build internal models of reality; short-cut computations may be just as good at choosing actions. In complex domains such as 3-D spatial cognition, which underpins many complex choices of action, the RE implies that brains build internal models of the animal's surroundings; and that the models are constrained to be true to external reality.

**Keywords**:  Requirement Equation; Bayesian cognition; short-cut cognition; internal models of reality.

## 1. Introduction

There is a debate between Bayesian model-based approaches to cognition, in which the brain constructs internal models of external states of affairs, and enactivist approaches, in which there are no internal models, but brains use simpler heuristics from sense data. This paper derives a criterion to decide, for any domain of cognition, which of these two approaches is appropriate.

I derive a a mathematical statement of the computation a brain is required to make (or more precisely, to choose actions **as if** it had made this computation) to give its owner the greatest possible fitness. This statement is the **Requirement Equation** (**RE**) and it defines mathematically what a brain is required to do for best fitness.

The Requirement Equation resembles Bayes' theorem, applied to states of the animal's habitat; but is Bayes' theorem; it has extra terms depending on fitness payoffs. The terms in the equation depend on things outside the animal's brain; they depend on physics, ecology and biology in the habitat.

The Requirement equation defines what computations brains need to do, not how they do them. Computing the RE needs no hypotheses about how brains work.

Consequences of the Requirement Equation include:

- Bayesian-like cognition gives the greatest possible fitness – providing a theoretical justification for Bayesian models of cognition.
- When there are few available actions, and the choice of actions has small information content, it may not be necessary to construct internal models of reality; short-cut computations may suffice.
- For animals with complex sense data and complex actions, internal models of external reality are needed for near-optimal cognition.
- The brain's internal model used in any domain must obey the constraints of the domain – such as the laws of 3-D geometry and kinematics, in spatial cognition.
- The full computations required for Bayesian optimal cognition are usually not feasible in brains; so brains must make simpler approximate calculations.

- The maximum possible fitness cannot be exceeded, no matter how complex or capable the brain is. Most animal brains may come close to this upper limit of fitness; more powerful brains would not give greater fitness.
- Brains evolve slowly to come closer to the optimum performance of the Requirement Equation. They come closest in respect of universal domain constraints, which have been true for all evolutionary time.

The requirement equation can be used to analyse cognition in domains including:

- Navigation and foraging
- Three-dimensional spatial cognition
- Multi-sensory integration
- Control of physical movements
- Learning
- Planning

Some applications are described in related papers [Worden 2024a, 2024b].

## 2. The Requirement for a Brain

This section derives a mathematical statement of the requirement for a brain – a statement of what brains need to do, in order to help their owners survive. The statement is independent of how brains compute (it does not assume the computation is done by neurons). It leads to a in section 3 requirement equation, which defines computably what brains are required to do for greatest fitness, rather than how they do it.

The derivation uses evolutionary modelling, and shares some assumptions with other evolutionary models [Maynard Smith 1982]:

- Evolution progressively refines the phenotypes of animals to improve their Darwinian fitness – which is the average number of copies of an animal's genotype passed on to the next generation
- At any time in its life, an animal has an expected Darwinian fitness, which depends on its internal state and the external state of the world.
- An animal's life consists of a sequence of encounters, during which it needs to choose actions to maximise the fitness payoff – which is the change in its expected Darwinian fitness from the encounter. The purpose of its brain is to choose the actions, depending on its sense data.
- Cognition can be approximately separated into a number of domains, with the cognitive goal of maximizing the fitness payoff in encounter in each domain.

I shall not discuss in detail how domains can be delineated, or the approximations that are needed to treat any domain on its own, in isolation from other domains. The analysis requires sensible approximations to apply the theory, to know how to combine different domains, and so on.

For each encounter, the input to the brain is a set of sense data (denoted by d), and the output of the brain is a choice of an action (denoted by a), which the animal does. These may depend on the existing state of affairs (denoted by s), in ways which will be defined. The state s includes everything external to the animal, and includes the internal state of its body.

To analyse what brains are required to do, we go a little beyond the 'black box' model of the brain (in which brains have input sense data d, and produce as output a choice of actions a) to a 'gray box' model which is shown in figure 1.

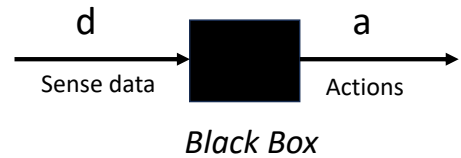

*Black Box*

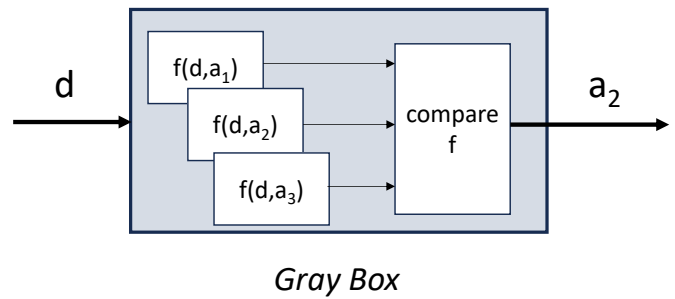

*Gray Box*

*Figure 1: Black Box and Gray Box models of a brain.*

The black box model makes no assumptions about how brains compute the actions a from sense data d; any computation is allowed. The gray box model goes a little further, assuming that the choice of actions is made 'as if' the following takes place:

- There is a set of possible actions $a_1, a_2,…$
- For each possible action, the brain computes a positive real function $f(d,a_1), f(d,a_2),..$ (these may be computed in parallel, or in any order)
- The brain compares all the $f(d,a_1), f(d,a_2),..$ and chooses the $a_i$ with the largest f

Compared with the black box brain, the gray box model makes only minimal assumptions. It assumes only that a set of possible actions is considered, and that one action is chosen by some criterion.

The animal's Darwinian fitness in state s will be denoted by G(s), which is also called the 'genetic payoff' – the number of copies of its genotype it is expected to pass on to the next generation.

The sense data d that the animal receives at any instant is the aggregate of its sense data of all modalities – its whole visual field, the sounds it hears, the configuration of its limbs, its internal sensations, and so on. As an encounter may have extended duration, the configuration of sense data d over an encounter is an aggregate of the configurations at each instant. The set of all possible sense data configurations may be a large set. The same goes for states of affairs, and for choices of action.

I use the lower-case letters s, d, a to denote individual states of affairs, configurations of sense data, and choices of action; and the corresponding upper-case letters to denote the sets of all possible states of affairs, and so on. P(s) denotes the probability of states, and so on. It then follows that:

$$\sum_S P(s) = 1 \qquad (1)$$

(in any encounter, some state of affairs must happen)

$$\sum_D P(d) = 1 \qquad (2)$$

(the brain must receive some configuration of sense data)

There is a joint probability P(s,d) of some state of affairs and some sense data occurring in the same encounter, and a conditional probability P(d|s), related to the joint probability in the usual way. Some of the variables may be continuous, in which case the P denote probability densities, and the sums are integrals over the densities.

In each encounter, the animal does not know which state of affairs holds; it only knows its sense data d. To the animal, states are hidden variables. Even its own internal state is revealed to its brain only as sense data. Based on the sense data it computes which action a to take. This is formalized as in figure 1: Given sense data d, the animal's brain computes some positive real function f(d,a) for all possible actions a, and chooses the action a which gives the largest f.

This formalization captures the idea that a brain may need to consider all possible actions a, in order to choose one action; it should not leave any action out of consideration. Because the set of all actions is typically a large set, it is not possible for a brain to examine all actions; we assume that brains have heuristics to examine only sensible actions; for all other actions, the effect of the brain is to set f(d,a) = 0.

Suppose that in state of affairs s, the animal receives sense data d, and chooses action a. Taking this action leads to an ensuing state s', which depends only on the earlier state of affairs and the action. Different states s' occur with conditional probabilities P(s'|s, a), such that for any s and a:

$$\sum_{s'} P(s'|s, a) = 1 \qquad (3)$$

(some next state must ensue, with probability 1)

Depending on the ensuing state, the animal receives a value v(s,s'), which is related to the animal's expected genetic payoff G in the two states:

$$v(s, s') = c_v[G(s') - G(s)] \qquad (4)$$

$c_v$ is a positive constant 'exchange rate' for v depending on the domain; the task of the brain in the domain is to choose an action a which makes the expected value v as large as possible (equivalently, makes G(s') as large as possible). For instance, in some feeding or foraging domain, v is related the amount of food obtained (not necessarily monotonically); or in some predator avoidance domain, v depends on the probability of becoming prey. Different domains have different currencies. The currencies are all inter-convertible into something related to the total Darwinian fitness G.

These terms in the meta-model of cognition are shown in figure 2:

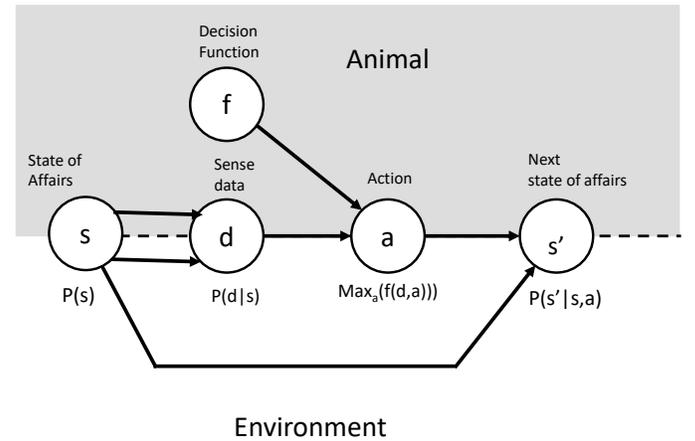

*Figure 2: meta-model of what brains are required to do, in one encounter in one domain*

The terms are:

1. In some domain of cognition, in an encounter of limited duration, states of affairs s occur with prior probability P(s)
2. States have variables relating both to the Animal's internal state (upper left arrow) and the external state (lower left arrow)
3. These states give rise to sense data d, with probability P(d|s). d depends on both internal and external aspects of the state.
4. The animal's brain acts as if it calculates a function f(d,a) for all possible actions a in a mutually exclusive set.
5. The animal does the action a which gives the largest value of f(d,a)
6. This leads to an ensuing state s', with probability P(s'|s, a)

7. This state gives a value v(s,s'), where v is measured in some currency which relates linearly to the animal's expected Darwinian fitness

When s, d and so on are described by continuous variables, finite probabilities P are probability densities in these variables, and the sums are integrals.

Then, the evolution of brains is a process of progressive refinement of the functions f(d,a) used in each domain, by random variation and selection, to increase the expected values v, and so to increase the expected number of offspring. The requirement for a brain is to compute some function f(d,a) (or rather, choose actions a 'as if' it had computed the f); and natural selection acts to make the function f progressively fitter.

This requirement for a brain is general. f(d,a) can have any functional form, and can be computed in any way. It is a statement about cognition at Marr's [1982] Level 1, which he called the 'computational' level, but which can be called a requirement level. It is about what brains are required to do, not about how they do it.

That completes the general description of the requirement for a brain in any domain. It is (for given sense data d) to compute some decision function f(d,a), as defined above, for all possible actions a, and then choose the action with largest expected value f.

## 3. Deriving the Requirement Equation

The Requirement Equation defines the best possible function f(d,a) which a brain needs to compute – or to choose actions (a) as if it has computed. It defines a brain which gives the greatest possible fitness.

Suppose that an animal brain uses some function f(d,a) to choose an action in each encounter. We can then compute the mean expected value received by the animal over all encounters, as follows (from figure 2):

- Different states of affairs s occur with probabilities P(s).
- For any state s, the animal may receive different sense data with probability P(s,d) = P(s)*P(d|s)
- For any sense data d, the animal computes f(d,a) for all a, and chooses the action $a_{max}$ with largest positive f.
- For any action a, states s' ensue with probability P(s'|s, a)
- The ensuing state has a value v(s,s') to the animal.

We can put these factors together to compute the expected mean value W to the animal of all encounters:

a(d) = the a which makes f(d,a) the largest    (5)

$$W = \sum_S P(s) \sum_D P(d|s) \sum_{s'} P(s'|s, a(d))v(s,s')$$

(6)

The derivation of equation (6) is just probability theory, applied to states of affairs, the resulting sense data, and ensuing states depending on actions taken by the animal.

Rearrange the sums to express the average value as a sum over sense data d:

$$W = \sum_D \sum_S P(s)P(d|s) \sum_{s'} P(s'|s, a(d))v(s,s')$$
$$= \sum_D J(d) \qquad (7)$$

where

$$J(d) = \sum_S P(s)P(d|s) \sum_{s'} P(s'|s, a(d))v(s,s')$$
(8)

W is the mean value to the animal from all encounters – the average fitness payoff in the domain, from any encounter. J(d) partitions W according to the sense data the animal receives – so that W is a sum or integral of J over all d, as n equation 7.

J depends on a(d), which is the action the animal chooses, depending on its sense data. So J depends on the animal's brain, through its decision function f(d,a). Now consider the case where the animal does not use its sense data, but chooses the same action (a) every time (a does not depend on d):

$$H(d, a) = \sum_S P(s)P(d|s) \sum_{s'} P(s'|s, a)v(s,s') \quad (9)$$

The sum of H(d,a) over d is also a value W – the mean value to the animal from all encounters, for a 'no-brainer' animal which always chooses the same action a. H does not depend on the animal's brain in any way – but, as can be seen from equation (9), H depends only on probabilities of states of affairs, sense data, and ensuing states, and on the fitness payoffs of those states. It depends on the physics, ecology, and biology of the animal's habitat. The function H(d,a), which depends only on things outside the brain, is used to analyse what the brain needs to do.

Now we switch to a special case, to see clearly what is happening. The special case is as follows:

- The sense data d consists of only one real variable, which can have values between 0.0 and 1.0.
- The animal only has a discrete choice of two actions, which are labelled 1 and 2.

In this case, H(d,a) consists of two functions $H_1(d)$ and $H_2(d)$, each defined over the range of d from 0.0 to 1.0, by equation (9) above. So:

$$H_1(d) = \sum_S P(s)P(d|s) \sum_{s'} P(s'|s, 1)v(s,s')$$

(10)

and similarly for $H_2$. Figure 3 shows a possible $H_1$ and $H_2$, plotted as functions of d:

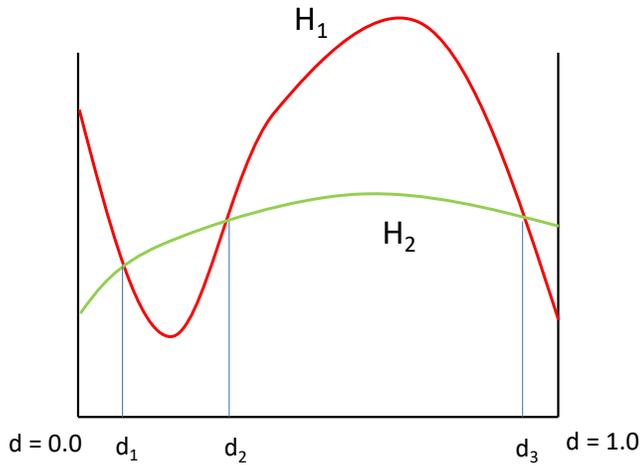

*Figure 3: Illustration of the choice of actions for an animal which can choose one of two discrete actions, using sense data d which is a single real variable in the range 0.0 to 1.0.*

If the animal always chooses action 1 (independent of its sense data d), then the average payoff W is the integral under the red curve of $H_1$ (this is the sum over all possible d). If it always chooses action 2, W is the integral under the green curve of $H_2$. In each case, W is a mean payoff for a no-brainer animal, which does not use its sense data.

The animal can do better than this, if it uses its sense data. Suppose it makes the following choices:

- For d between 0.0 and $d_1$ (where the $H_1$ curve is higher) it chooses action 1
- For d between $d_1$ and $d_2$ (where the $H_2$ curve is higher) it chooses action 2
- For d between $d_2$ and $d_3$ it chooses action 1.
- For d between $d_3$ and 1.0, it chooses action 2.

If it makes these choices, the average value W is an integral over d from 0.0 to 1.0, for each d using $H_1$ or $H_2$ – whichever curve is higher. You can see that this set of decision thresholds $d_1…d_3$ gives the best possible choice of actions – no other choice can give a higher value[1] of W.

Therefore the best possible decision function f(d,a) is given by f(d,a) = c(d)H(d,a), where c is any function of d (different choices of c(d) all give the largest f at the same a for any d – and so give the same choices of actions a)

Now move back from this special case to the general case:

- If there are not just two possible actions, but some larger number N of actions, the diagram of figure 3 will have N different curves; but the best action for any d is still to choose the highest curve at that d.

- A continuum of possible actions is the limit of a very large discrete set of actions.
- If the sense data d consists of not just one real variable, but of two or more (M) real variables, the H curves become shapes in M dimensions. The best action a is still to choose the highest H at any point d in this space
- If the sense data d include some discrete variables, the analysis proceeds similarly for each discrete value in d.

In the general case, the best choice of decision function is f(d,a) = c(d)H(d,a), where c is any positive function of d, and H depends only on events outside the brain. No other decision function can give greater fitness.

The result is that using the decision function

$$f(d,a) = \sum_s P(s)P(d|s) \sum_{s'} P(s'|s,a)v(s,s') \quad (11)$$

gives the greatest possible fitness (by generalising the special case described above, to more possible actions and more sense data)

Equation (11) is the result of the paper, and it is called the **Requirement Equation (RE).** The best decision function f (which gives the greatest possible fitness) results from the brain computing the decision function f(d,a) for all possible actions a, and then choosing the action which gives the largest result.

A simpler form of the requirement equation is given later in equation (16).

Equation (11) resembles Bayes' theorem, applied to the 'hidden' states s and sense data d, but it is not the same formula, because of the presence of actions a, ensuing states s', and values v.

The derivation shows that for any animal, in any domain and habitat, choosing actions by equation (11) gives the greatest possible fitness – the greatest possible average value v across all circumstances, in whatever value currency for v is appropriate for the domain.

The Requirement Equation can be understood using the concept of no-brainer animals, as above. Suppose you have a large set of no-brainer animals, each of which can only do one action a; there is one animal which does each action. Depending on the sense data d, you are allowed to switch to whichever animal does best for that set of sense data. The resulting fitness is that of the RE; it is better than the fitness of any one animal, and it is the fittest that can be achieved.

Equation (11) is a requirement for a brain; a statement of what a brain needs to do, to achieve the best possible fitness. The terms in equation (11) depend on factors outside the animal's brain – the physics, biology and ecology of the animal, its sense organs, its possible actions, and its habitat. They can be modelled with as much precision as needed, provided you know enough of the biology and ecology. Then, in most cases of interest equation (11) can be computed[2] (sometimes by brute force numerical integration). The RE is a computable statement about what brains need to do.

---

[1] If any of the decision thresholds such as $d_2$ is altered by a small amount ε, W decreases by an amount proportional to $ε^2$.

## 4. Short-Cut Cognition

As it stands, the Requirement Equation (11) is too expensive to compute in a brain. To compute it, not only must the brain compute f(d,a) for all actions a; but also, the computation for each action a requires a sum over all possible states s. Computing equation 11 is expensive in four ways:

1. f(d,a) needs to be computed for all possible actions a, to select the largest f(d,a).
2. Computing f(d,a) requires a sum over all possible states s, weighted by their prior probability
3. For any action, it requires a computation of all possible ensuing states s'.
4. In some cases, the calculation of the value requires a lookahead to a branching tree of future states.

We would not expect brains to make this quadruply expensive computation. A near-optimal brain delivers a choice of actions **as if** it had made the computation. From the point of view of the animal's fitness, the brain is a black box. It does not matter how it computes the choice of action, as long as it gets the fittest answer, or the nearly fittest choice of action – as if it had computed all the f(d,a).

In some cases, it is possible to compute the best action by a short-cut computation, which bears no resemblance to the requirement equation, and involves no internal representation of states of affairs.

An example is the case which was used in the previous section to explain the derivation of the requirement equation – where the animal has only two possible actions, and only one sense datum d in the range 0..1.

To compute the requirement equation, this animal would compute both the functions $H_1$ and $H_2$ of equation (10), and compare the two values to decide which action to take. Calculating each H is a complex double sum over possible states of affairs. That is not necessary. The short-cut computation is just to compare its sense data d with the thresholds $d_1$, $d_2$ and $d_3$ of figure 3, and to choose its action accordingly. Testing d against three inequalities is a much simpler computation – an economical and effective short-cut.

Similar short cuts may be possible with more complex sense data, as long as the choice of actions is simple. It may be possible to find regions in the space of sense data, where certain choices of action are the best. As long as the choice of action is highly restricted, and has a small information content, there may be a feasible short cut computation. In these cases, evolution would not waste resources on a brain doing a more complex computation.

One way to see this is to think of the brain as an information channel, whose inputs are sense data, and whose only output is a choice of actions. If the output has small information content, it may be possible to compute it without any complex computation. The brain is then an enactivist brain [Hutto & Myin 2012, Galloway 2017], which simply responds to sense data.

Simple choices of action are a necessary condition for short-cut cognition; but they are not a sufficient condition. Even for simple choices of action, a simple short-cut computation may not exist.

For instance, animals have many binary choices of action: eat/do not eat; hide/flee; sleep/wake, and so on. In each case, the choice of action has small information content, but there may be no simple way to compute it. These binary choices usually depend on many factors – including, for instance, where the animal is. Is it near home, or in its burrow? Is it in a safe area, or near a place where it has seen predators? These decisions all depend on the animal knowing where it is. The best way to know where it is, is for the animal to build an internal cognitive map of its habitat, using many kinds of sense data. A cognitive map is the kind of internal model of reality which the requirement equation (11) leads to; building it is not a short cut.

Short-cut cognition is shown in figure 4:

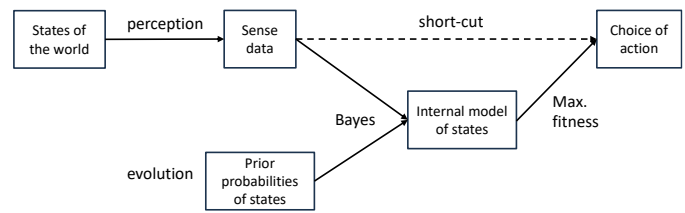

*Figure 4: Short-cut cognition*

The computation suggested by the requirement equation is shown by the solid arrows, and described further in the next section. Here, sense data are combined with prior probabilities of states (which have evolved into the animal's brain over many generations) to compute a maximum likelihood model of the states of the world. This model is then used to evaluate all the possible actions, to find the action with the maximum expected fitness payoff. Short-cut cognition is a direct computation from sense data to actions (dashed arrow) - a heuristic which requires no internal model of reality.

---

[2] As is described in [Worden 2024a], there are issues in computing the RE for the case of multi-step planning. For instance, while the RE could be used formally to describe the requirement for a perfect chess player, the resulting form is not computable. Hard planning problems remain hard planning problems.

Short-cut cognition is not appropriate when several choices of action depend on the same aspect of reality, such as where the animal is geographically. This is shown in figure 5.

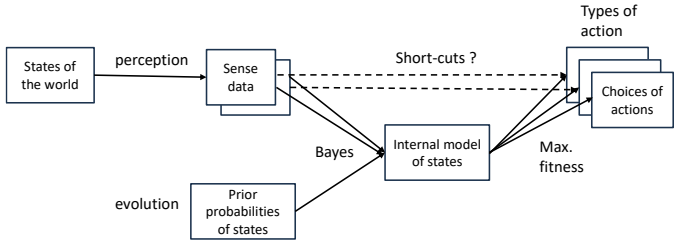

*Figure 5: Multiple types of action depend on the same aspect of reality. The picture also shows that many types of sense data contribute. The short-cut calculations (dashed arrows) are not feasible.*

Here, there are several different choices – for different types of action, which can be chosen independently. They may be binary choices such as 'sleep/wake' or more information-rich choices like 'in which direction should I go?'. They all depend on the same aspect of reality, such as 'where am I?'

It is very unlikely for there to be good short-cut computations for all the different choices of action. For one or more of the choices, some internal model of reality – such as a cognitive map – is required. Therefore the animal's brain is required to build this model. If a model has to be built, it can be used for all the different choices of action. Because the model describes an aspect of reality that many choices of action depend on, and uses many kinds of sense data to build it, it is likely to give a better choice for all the actions – better than direct short cuts from sense data.

So in cases where animals have complex sense data, and where many choices of action depend on the same aspect of reality, short-cut cognition is unlikely to be used. In these cases, internal models of reality are likely to be used.

## 5. Bayesian Models of Reality

To see how and when Bayesian models of reality are entailed by the requirement equation (11), note that the functions f(d,a) can be multiplied by any positive function of the sense data d, and that will not alter the choice of action a which gives the largest f. So define

$$f'(d,a) = \frac{f(d,a)}{P(d)} \qquad (12)$$

And use Bayes' theorem

$$P(s|d) = \frac{P(s)P(d|s)}{P(d)} \qquad (13)$$

to derive from equation (11):

$$f'(d,a) = \sum_s P(s|d) \sum_{s'} P(s'|s,a)v(s,s') \qquad (14)$$

Define

$$W(s,a) = \sum_{s'} P(s'|s,a)v(s,s') \qquad (15)$$

To give

$$f'(d,a) = \sum_s P(s|d)W(s,a) \qquad (16)$$

This is an alternative form of the Requirement Equation. The interpretation of W(s,a) is simple; W is the average fitness payoff of taking action a in state s. The best choice of action is to choose the a which gives the largest f'(d,a).

It is useful to regard equation (16) as follows: the space of possible states s is a vector space (often of very high dimension). P(s|d) defines a vector **d** in that space. For a mutually exclusive set of actions $a_i$, each W(s,$a_i$) is a vector **a**$_i$ in the same space. f'(d,a) is the inner product of these two vectors:

$$f'(d,a_i) = \mathbf{d}.\mathbf{a}_i \qquad (17)$$

Then the fittest choice of action a is the choose the action $a_i$ which gives the largest inner product of $a_i$ and d.

This can be seen by an analogy to a discrete Fourier transform. Suppose the state s is defined by a real variable in the range 0…100. The probability density P(d|s) = **d** is a positive function in this range. Suppose it has Fourier components $d_n$. These are components of a vector in a Hilbert space, as are the components $a_{in}$ of **a**$_i$. f'(d,$a_i$) is the inner product.

Up to this point, we have made no approximations; equation (16) or (17) is still a precise statement of the fittest possible cognition. The probabilities in it are objective facts about the world, the habitat, and the biology of the species. For instance, P(s) in equation (11) is an objective probability (a measurable frequency) of states of affairs in the animal's habitat. If the state s depends on locations of things, P(s) has zero probability density for states that do not obey Euclidean geometry.

Now make approximations. Assume that:

1. Some dimensions of the state s are continuous variables; there may be a large number of variables, and they are denoted by $x_i$, i = 1..N. Probabilities P are probability densities in the $x_i$, and the sums are integrals
2. Given the animal's sense data d, there is one most likely state s = $s_{max}$, with $x_i$ = $x_{i0}$ for i = 1..N.
3. The probability density P(s|d) has an approximately Gaussian distribution around the maximum in all the variables.

4. The animal has fairly precise sense data d, so that in many of the variables $x_i$, the Gaussian distributions ae fairly narrow.

The dependence of P(s|d) on one of the variables $x_i$ is

$$P(s|d) \sim exp\left(-\frac{(x_i - x_{i0})^2}{\sigma_i^2}\right) \quad (18)$$

Point 4 states an assumption that the standard deviation $\sigma_i$ is small, giving a narrow Gaussian peak.

How does this affect the choice of fittest action through equation (16)? If P(s|d) has a narrow Gaussian peak, then the payoff functions W(s,a) do not vary much across the peak – giving, to a good approximation:

$$f'(d,a) = W(s_{max}, a) \int P(s|d)$$
$$= W(s_{max}, a) \quad (19)$$

This is the Bayesian maximum likelihood approximation. The fittest choice of actions is got by finding the maximum likelihood state $s_{max}$, evaluating the fitness payoff W for each action in that state, and choosing the action with the largest payoff. This is a simpler computation than the full integral of (16), so the assumptions (1) –(4) imply that animals often use this computation.

Having derived the maximum likelihood approximation from first principles, we can analyse where it may go wrong, and how it may go wrong. It can go wrong when the Gaussian peaks are not very narrow, and the payoff functions W vary appreciably across the peaks.

The maximum likelihood approximation is good if:

- The probability density P(s|d) is a narrow peak in the variables of s, and is a fairly symmetric function on either side of the peak.
- The fitness payoff functions W(s,a) are slowly varying functions of s

Conversely, it is not a good approximation if:

- The probability density P(s|d) is a broad peak in the variables of s, or is asymmetric on either side of the peak.
- The payoff functions W(s,a) vary rapidly in s

The maximum likelihood approximation is a good one if the sense data are highly discriminating, and if they effectively narrow down the range of possible states. Because sense organs have evolved to give very discriminating sense data, it is often a good approximation.

There are two important cases where there are many different types of action (which can be chosen independently) which all depend on the same aspect of the external world. These cases are:

1. Many choices of action depend on where the animal is geographically within its habitat. These decisions depend on the animal knowing its most likely location, in a two-dimensional cognitive map of its territory.
2. The control of physical actions depends on knowing the three-dimensional positions and movements of things close to the animal (including its own limbs). This requires an internal 3-D model of local space.

In these cases, because many actions depend on the same aspect of reality, short-cut cognition is not expected to be the fittest possible, and brains are expected to build Bayesian internal models of external reality.

## 6. Constraints on Models of Reality

I have shown that in two important domains (a 2-D cognitive map, and a 3-D model of local space), both of which underpin many diverse choices of action, brains need to build internal models of reality.

There is a question of how much the models need to be truthful to reality, or veridical. [Hoffman 2009, Hoffman, Singh & Prakash 2015] have argued that the models in the brain built from perception are merely interfaces to the external world, and they are not required to be veridical models. They go further, to assert that the models are not veridical.

This section describes some constraints on how far the internal models can depart from reality, in the two domains above (cognitive maps, and 3-D spatial models). It shows that in those domains, the models cannot depart far from reality.

The external states of the world, on which the animal's fitness depends, obey the laws of geometry, kinematics and physics:

a) The geographic locations of important things in the animal's habitat – such as its burrow, or food sources – obey the constraints of 2-D Euclidean geometry.
b) Many of the objects in the animal's habitat do not move.
c) The locations of objects close to the animal obey the laws of three-dimensional Euclidean geometry
d) Movements of objects obey kinematic constraints, relating positions to velocities. For instance, if an object moves towards the animal, the distance to it reduces.
e) Movements of objects obey physical constraints – they tend to move in straight lines or fall downwards; heavy objects have inertia, some surfaces resist pressure, and so on.

Any animal whose choices of physical actions flew in the face of these constraints would be much less fit because of it – like Wile E Coyote, always falling out of the sky:

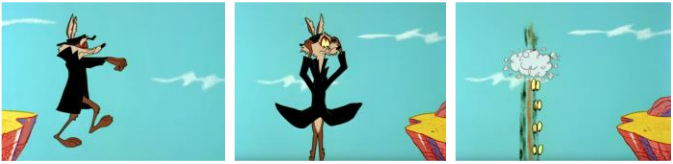

*Figure 6: What happens to an animal, which does not control its movements with an internal model of 3-D space that obeys the laws of geometry and physics*

These physical constraints have been true for all evolutionary time. They govern the prior probabilities of states of the world; states that violate the constraints do not occur, so have zero prior probability density. Those probability constraints have evolved into the brains of animals, over all evolutionary time. We would expect brains to reflect those constraints very precisely, after half a billion years of evolution – as precisely as brains can compute.

In this first respect, animals' internal models of reality, built from sense data, are constrained to be veridical. The models are built using the real prior probabilities of the world – which obey the laws of geometry, kinematics and physics. That severely limits how far the models can be distorted or untruthful.

There are further constraints. Consider different types of action a, b, c…, all of which depend on the same set of states s. For action type a, there are mutually exclusive actions $a_1$, $a_2$…, with fitness payoffs $W(s,a_i)$, and similarly for action type b with fitness payoffs $W(s,b_i)$. Actions of types a and b can be chosen independently.

As well as discrete choices of action like sleep/wake or fight/flee, there are many choices of continuously variable actions. To choose one of these actions is to fix a small set of real numbers, making choices like:

- In what direction should I travel to go home?
- What muscle force should I exert to stand up?
- In what direction should I move a limb to grasp an object?
- How should I compensate my balance when pushed?

There are many continuously variable choices of action, and they are needed at all moments of the day. We denote the continuous action variables by $y_i$. Typically they result in graded commands to muscles.

The state s in either of the two spatial models is described by a set of real variables $x_i$ defining the locations of things. For instance, in a cognitive map with only three landmarks and a defined overall direction, by Galilean invariance there are four independent variables $x_1…x_4$. This does not depend on whether the map uses Cartesian coordinates, polar coordinates, or some other coordinate system.

Suppose that in some dimension $x_i$, the likelihood $P(s|d) = P(x)$ is not a narrow Gaussian; it is a broader function of any dimension x, and may be asymmetric. There is a mean value $x_{mean}$ of the variable x in the likelihood distribution $P(x)$, as well as the maximum likelihood value $x_{max}$. $x_{mean}$ and $x_{max}$ in general different. Any of the fitness functions $W(x_i, y_i)$ may vary appreciably across a peak of large likelihood.

The functions $W(x_i, y_i)$ do not depend on events inside the animal's brain. They only depend on how the events following its actions affect its survival. The W could in principle be measured by observing animals in the wild – what they do, and how well they survive. We make the following assumptions about the W:

- For give state variables $x_i$, the fittest choice of actions $y_i$ depend on the $x_j$, through a functional form $y_{fit} = q(x_j)$.
- If the animal chooses different action $y_i$, the fitness function decreases quadratically away from the fittest value, in a region of small $(y_i − y_{fit})$, where the function q is approximately linear. This is written as $W(x_j, y_i) = W_0 − c(y_i − y_{fit})^2$, where $W_0$ and c are positive constants.

These are weak assumptions about the fitness functions W. They are likely to hold for many physical actions, like the examples cited above.

With these assumptions, it can be shown that $y_{fit} = q(x_{mean})$. [Mamassian, Landy & Maloney 2002]. This follows from equation (16) because the variance of a function about a value is minimized when the value equals the mean; and ($W_0 − W$) has the mathematical form of a variance.

So even when the posterior probabilities $P(s) = P(x_i)$ are broad, non-Gaussian functions of the state variables $x_i$, there is still a best Bayesian model $s = s(x_{imean})$, which can be used for many different continuous-variable choices of action, by choosing $y_{fit} = q(x_{mean})$. There is still a Bayesian best choice of actions, using the same model for all actions. It uses the means of likelihood peaks, and is not exactly the maximum likelihood model.

This result holds for modest departures of the likelihood functions from a narrow Gaussian approximation. It does not hold for extreme departures, such as a bimodal distribution in some $x_i$. In these cases, the approximation that W is quadratic may not hold as far as to span the two peaks, and the brain must choose one peak or the other. This accounts for the switching of ambiguous figures such as the Necker Cube; the brain takes one interpretation or the other, not in between.

The best Bayesian model to use for many choices of action is not a maximum likelihood model, or a mean model; it is to find maximum likelihood peaks, and to find mean values within the peaks. The key point is that this model gives the best choice of actions for many different types of physical action.

There is a third sense in which Bayesian models of space (2-dimensional or 3 dimensional) are constrained to be truthful. The state s of such a model at one moment in time, defined by sense data d, can be regarded as a vector **d** in an N-dimensional space of the variables $x_i$, where N is typically large. Over time, the state point moves, and many different choices of action need to be made.

Equation (17) is an inner product **d.a**$_i$ of the state vector **d** (describing the posterior probabilities of states, given sense data d) with a fitness vector **a**$_i$ for the choice of action $a_i$. $a_i$ is chosen to maximise the inner product. **d** expresses knowledge of the state s from sense data – P(s|d). For diverse types of action a, b,…, there are many different vectors **a**$_i$, **b**$_j$ and so on. Each different choice of action constrains the state vector **d** in one or more directions; if the component of **d** in that direction is inaccurate, poor fitness follows. Mistakes are expensive. The many different choices of action measure the state vector **d** along many different dimensions. To make those choices well requires the state vector **d** = P(s|d) to be accurate (i.e. to resemble the real world) in all those dimensions. Many different types of action fix many independent components of the state vector – so that to choose all the actions well, it needs to be a veridical state vector.

In summary, there are three sets of constraints on the internal models of 2-D and 3-D space:

1. The models must obey the constraints of geometry, kinematics and physics – which have had a very long time to evolve into the prior probabilities used animal brains, and so are accurately reflected in brains.
2. Many types of continuously variable actions can be computed from the same state model – but they need to be computed from accurate mean values of the state, close to real-world values, in order to be as fit as possible.
3. Many different types of action fix many independent components of the state in a multi-dimensional state space, forcing the state vector to be true to reality.

The examples used by [Hoffman 2009, Hoffman, Singh & Prakash 2015] to motivate the interface theory of perception do not relate to the control of physical movement. They are about categorization, to choose one action out of few alternatives, such as:

- a jewel beetle mating choice
- a herring gull chick feeding choice
- a frog eating only moving things
- a dragonfly choosing to lay eggs

These categorical choices have low information content, and do not require veridical internal models of the world (they can use short-cut cognition); but the physical control of motion involves many graded choices, which depend on a truthful internal model of space. All these animals use veridical models to control movement.

The interface theory of perception says that because internal models of reality have evolved just to maximise fitness, they need not be veridical. I have considered two internal models of reality – a two-dimensional geographical model (or cognitive map), and a three-dimensional spatial model. These models need to be true to the external world in all these respects. So these models are highly constrained and have little freedom to be distorted or non-veridical.

## 7. Is the Bayesian Brain a Hypothesis?

The 'Bayesian Brain' hypothesis [Mamassian et al 2002; Knill & Pouget 2003] is that brains use Bayes' theorem to compute the most likely state of affairs consistent with their sense data. This hypothesis has been very successful in comparison with data. It is usually treated as a hypothesis to be tested. This paper suggests it is more than that.

The requirement equation (11) and (16) was derived from general assumptions about evolution and the role of cognition in achieving fitness. This derivation shows that the Bayesian Brain is a provable property of brains, and can be proved from general assumptions about evolution. Brains evolve to maximise fitness payoffs to the animal. For complex choices of action, this requires Bayesian cognition. This does not imply that all brains are always Bayesian, but it implies that they evolve to give a choice of actions which keeps coming closer to the Bayesian answer – because that is the answer that gives greatest fitness.

Because equations (11) and (16) define the fittest cognition possible, it is not possible to be fitter than that (without improvements to sense organs or available actions). The requirement equation defines the best possible brain, and there can be no better brain. If the choice of actions resulting from the requirement equation – or a good approximation to it – can be computed in any way, adding further computing power would serve no purpose. So it may be that many animals with limited intelligence have as much intelligence as they need; bigger brains would be a waste of resources. This is consistent with the fact that most species use only a small part of their energy budget in their brains – because to use any more would not give them added fitness.

## 8. The Evolution of Brains

The Requirement Equation gives the greatest possible fitness, only if the probabilities used to calculate the f(d,a) match the actual probabilities in the animal's habitat.

Improving the brain's approximation to f(d,a) means that over evolutionary time, the internal probabilities which are used to compute f (or as if to compute the f), converge towards the actual probabilities in the habitat (as far as the resource limits of the brain allow them to).

The probabilities in the brain will converge towards those in the habitat, if probabilities in the habitat are stable. If the probabilities in the habitat fluctuate over time, the probabilities in the brain will converge too slowly to catch up. There is a speed limit for evolution [Worden 1995b, 2022] which implies that the probabilities built genetically in the brain can only converge slowly to the probabilities in the habitat. Greater precision in the internally represented probabilities requires more genetically encoded information, and this information can only accumulate at a very limited rate – typically less than 1 bit per generation.

An important challenge to Bayesian cognition arises when the prior probabilities of events in the animal's habitat change so rapidly that their brains cannot evolve fast enough to incorporate those probabilities. In this case, the animal needs to learn about the probabilities of events during its short lifetime. Their ability to learn fast can be understood in a Bayesian framework, as is described in a related paper [Worden 2024a]

## 9. Discussion

This paper has derived a Requirement Equation[3], which defines an optimum form for cognition.

Consequences of the equation include:

- The equation gives the greatest possible fitness – providing a theoretical justification for Bayesian models of cognition.
- For animals with complex sense data and actions, internal models of external reality are needed for near-optimal cognition.
- The brain's internal model used in any domain must obey the constraints of the domain – such as the laws of 3-D geometry and kinematics, in spatial cognition.
- In other ways, the brain's internal models of 2-D space (maps) and 3-D local space are constrained to be veridical.
- The maximum possible fitness of the RE cannot be exceeded, no matter how complex or capable the brain is. Many animal brains may come close to this upper limit of fitness; more powerful brains would not make them fitter.
- Brains evolve slowly to come closer to the optimum performance of the Requirement Equation. They can come closest in respect of universal domain constraints, which have been true for all time.

There are further consequences of the equation. In a related paper [Worden 2024a] I discuss applications of the Requirement Equation, including learning, navigation in two dimensions, planning, and testing cognitive models such as the Free Energy Principle [Friston, Kilner & Harrison 2006; Friston 2010]. In [Worden 2024b] I describe how the requirement equation applies to three-dimensional spatial cognition, illustrated by bees and bats.

Acknowledgement: I thank Chris Fields for comments which have helped to clarify the treatment.

---

[3] The requirement equation was derived in [Worden 1995b].